# Airborne Ultrasonic Tactile Display Brain-computer Interface
## - A Small Robotic Arm Online Control Study -

Tomasz M. Rutkowski[*,1,2,▽], Hiromu Mori[1], Takumi Kodama[1], and Hiroyuki Shinoda[3]

[1]Department of Computer Science and Life Science Center of TARA, University of Tsukuba, Japan

[2]RIKEN Brain Science Institute, Wako-shi, Japan

[3]The University of Tokyo, Tokyo, Japan

e-mail: tomek@bci-lab.info ([▽]the corresponding author)

*Abstract* We report on an extended robot control application of a contact-less and airborne ultrasonic tactile display (AUTD) stimulus-based brain-computer interface (BCI) paradigm, which received last year The Annual BCI Research Award 2014. In the award winning human communication augmentation paradigm the six palm positions are used to evoke somatosensory brain responses, in order to define a novel contact-less tactile BCI. An example application of a small robot management is also presented in which the users control a small robot online.

## I. INTRODUCTION

The state-of-the-art BCI applications are typically based on mental visual or auditory, as well as motor imagery paradigms, which require extensive user training and good eyesight or hearing. In the recent years alternative solutions have been proposed to make use of a tactile (somatosensory) modality [1,2,3] to enhance brain-computer interfacing efficiency and to allow robotic control applications.

The concept reported in this abstract further extends the brain somatosensory channel by utilizing a contact-less stimulus generated with an airborne ultrasonic tactile display (AUTD) [4] in application to a small robot control. The rationale behind the use of the AUTD is that it allows for more hygienic, and avoiding skin ulcers occurrences in locked-in state patients (LIS), application due to its contact-less nature. This abstract reports the very encouraging results with AUTD-based BCI (autdBCI) to an online and a "mental effort-based-only" robotic arm control.

## II. METHODS

This abstract is an extension of the previously published by the authors, and awarded, multiuser study with the autdBCI [6]. In the current pilot study two male volunteer BCI users participated in the online robot control experiments. The users' mean age was 34.0, with a standard deviation of 14.14 years. The experiments were performed at the Life Science Center of TARA, University of Tsukuba, Japan. The online (real-time) EEG experiments were conducted in accordance with *the WMA Declaration of Helsinki - Ethical Principles for Medical Research Involving Human Subjects* and the procedures were approved and designed in agreement with the ethical committee guidelines of the Faculty of Engineering, Information and Systems at University of Tsukuba, Japan.

The AUTD stimulus generator produced an airborne vibrotactile stimulation of the human skin using a focused ultrasound technique [4,5]. The effect was achieved by generating an ultrasonic radiation static force produced by intense sound pressure amplitude due to a nonlinear acoustic phenomenon [4,5]. The radiation pressure deformed the surface of the skin on the fingers and palms, creating a vibrotactile sensation. An array of ultrasonic transducers mounted on the AUTD created the focused radiation pressure at an arbitrary focal point due to an implementation of the so-called phased array technique. The AUTD-based BCI [4,5] adhered to ultrasonic medical standards and did not exceed the permitted skin absorption levels. It was approximately 40 times below the limits for medical device standards in Japan. The effective vibrotactile sensation was set to 50 Hz to match with tactile skin receptors and a notch filter set for power line interference rejection applied to EEG brainwave signals [5,6]. The users were instructed to conduct simple object grasping and movement sequences using a six commands-based robotic arm as shown in Figure 1 (see also an online video from the experiment available at [7]). The EEG signals were captured with an EEG amplifier system g.USBamp by g.tec Medical Engineering GmbH, Austria, using eight active electrodes positioned over the parietal cortex. The ground electrode was attached to the forehead, and the reference to the left earlobe. No electromagnetic interference was observed from the AUTD operating with frequencies notch–filtered together with power line interference from the EEG. The EEG signals captured were processed online with a BCI2000–based application, using an in-house improved stepwise linear discriminant analysis (SWLDA) classifier with features drawn from $0 - 800$ ms ERP intervals decimated by a factor of 20. The stimulus length was of 100 ms and inter–stimulus–interval were set to 300 ms (shorter comparing the previous studies [5,6]). The number of epochs to average was set to three in online tests (15 in the first classifier training data collection run), which was also shorter comparing to our previous publications [5,6]. The EEG recording sampling rate was set at 512 Hz. The high- and low-pass filters were set at 0.1 Hz and 60 Hz, respectively. The notch filter to remove power line interference was set for a rejection band of $48 \sim 52$ Hz.

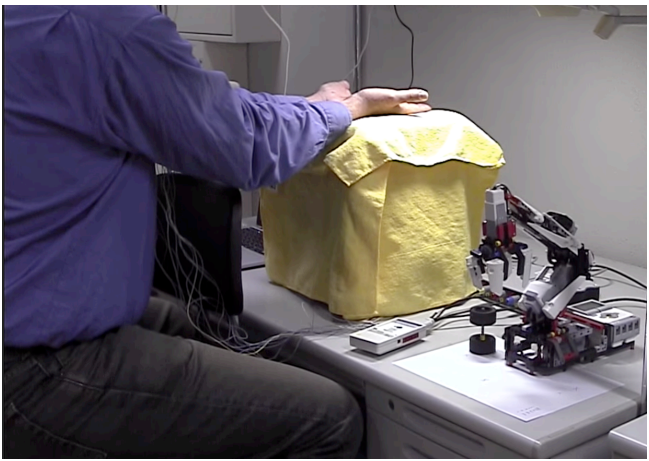

Figure 1. The autdBCI user with hands placed under the AUTD device and operating mentally the small six commands-based robot placed on the right side.

### III. Results and Conclusions

The theoretical chance level of the autdBCI-based robot arm operation was of 16.6%. The users managed to achieve successful scores after a short training by completing a simple six steps based small object picking and moving task as shown in Figure 1 or in an online video [7]. This case study demonstrates results obtained with a novel six-command-based autdBCI paradigm applied to the mental control of the robot.

The experiment results obtained in this study confirm the validity of the contact-less autdBCI for interactive robotic applications or human motor function augmentation. The results presented offer a step forward in the development of novel neurotechnology applications. Due to the still low number of tested users, the current paradigm obviously requires improvement and further testing. These requirements determine the major lines of study for the future research. However, even in its current form, the proposed autdBCI can be regarded as a practical solution for LIS patients (locked into their own bodies despite often intact cognitive functioning).

References

[1] G.R. Muller-Putz, R. Scherer, C. Neuper, and G. Pfurtscheller. Steady-state somatosensory evoked potentials: suitable brain signals for brain-computer interfaces? Neural Systems and Rehabilitation Engineering, IEEE Transactions on, 14(1):30–37, March 2006.

[2] A-M. Brouwer and J.B.F. Van Erp. A tactile P300 brain-computer interface. Frontiers in Neuroscience, 4(19), 2010.

[3] T.M. Rutkowski and H. Mori, "Tactile and bone-conduction auditory brain computer interface for vision and hearing impaired users," Journal of Neuroscience Methods, p. Available online 21 April 2014.

[4] T. Iwamoto, M. Tatezono, and H. Shinoda. Non-contact method for producing tactile sensation using airborne ultrasound. In Manuel Ferre, editor, Haptics: Perception, Devices and Scenarios, volume 5024 of Lecture Notes in Computer Science, pages 504–513. Springer Berlin Heidelberg, 2008.

[5] K. Hamada. Brain–computer interface using airborne ultrasound tactile display. Master thesis, The University of Tokyo, Tokyo, Japan, March 2014. In Japanese.

[6] K. Hamada, H. Mori, H. Shinoda, and T.M. Rutkowski, "Airborne ultrasonic tactile display brain-computer interface paradigm," in Proceedings of the 6th International Brain-Computer Interface Conference 2014 (G. Mueller-Putz, G. Bauernfeind, C. Brunner, D. Steyrl, S. Wriessnegger, and R. Scherer, eds.), pp. Article ID 018–1–4, Graz University of Technology Publishing House, 2014.

[7] T.M. Rutkowski, The autdBCI and a robot control (the winner project of The BCI Annual Research Award 2014);. Available from: http://youtu.be/JE29CMluBh0.